\documentclass[review]{elsarticle}

\usepackage{lineno,hyperref}
\usepackage{graphicx}
\usepackage{esvect}
\graphicspath{ {Figures/} }

\journal{Journal of Quantitative Spectroscopy and Radiative Transfer }

\usepackage{numcompress}

\begin{document}

\begin{frontmatter}

\title{Dual-Band Quasi-Coherent Radiative Thermal Source}

\author[UofA]{Ryan Starko-Bowes}
\author[Purdue]{Jin Dai}
\author[Purdue]{Ward Newman}
\author[UofA]{Sean Molesky}
\author[Beijing]{Limei Qi}
\author[Purdue]{Aman Satija}
\author[UofA]{Ying Tsui}
\author[UofA]{Manisha Gupta}
\author[UofA]{Robert Fedosejevs}
\author[UofA]{Sandipan Pramanik}
\author[Purdue]{Yi Xuan}

\author[Purdue]{Zubin Jacob\corref{mycorrespondingauthor}}

\cortext[mycorrespondingauthor]{Corresponding author}

\address[UofA]{University of Alberta, Edmonton, AB, Canada}
\address[Purdue]{Birck Nanotechnology Center, Purdue University, West Lafayette, IN, USA}
\address[Beijing]{School of Electronic Engineering, Beijing University of Posts and Telecommunications, Beijing, China}


\begin{abstract}
Thermal radiation from an unpatterned object is similar to that of a gray body. The thermal emission is insensitive to polarization, shows only Lambertian angular dependence, and is well modeled as the product of the blackbody distribution and a scalar emissivity over large frequency bands. Here, we design, fabricate and experimentally characterize the spectral, polarization, angular and temperature dependence of a microstructured SiC dual band thermal infrared source; achieving independent control of the frequency and polarization of thermal radiation in two spectral bands. The measured emission of the device in the Reststrahlen band (10.3-12.7~$\mu m$) selectively approaches that of a blackbody, peaking at an emissivity of 0.85 at $\lambda_x=11.75~\mu m$ and 0.81 at $\lambda_y=12.25~\mu m$. This effect arises due to the thermally excited phonon polaritons in silicon carbide. 
The control of thermal emission properties exhibited by the design is well suited for applications requiring infrared sources, gas or temperature sensors and nanoscale heat transfer. Our work paves the way for future silicon carbide based thermal metasurfaces.
\end{abstract}

\begin{keyword}
Thermal Radiation\sep SiC \sep Dual-Band Emission \sep Polarized Thermal Emission \sep Infrared Photonics \sep Selective Thermal Emitter \sep Phonon Polariton
\end{keyword}

\end{frontmatter}

\section{Introduction}
Thermally excited light is traditionally thought of as being incoherent, unpolarized and spectrally broad. These notions have been challenged in the past decade to overcome incoherent properties of thermal radiation sources. Many different approaches have been used to engineer the spectrum \cite{PadillaBlackbody,MoleskyTopologicalTrans,PendarkerThermGraph}, coherence \cite{Greffet2D,Greffet1D}, polarization \cite{Metal2DHan:13,Greffet2D} and directionality properties \cite{GreffetLaroche:1DW,Greffet1D,Greffet2D} of thermally emitted radiation. Proper design of ruled gratings in metals \cite{GreffetLaroche:1DW,ShalaevRefractPlasmon,ZHANG1DGrating} or polar dielectrics \cite{Greffet1D} have been successfully employed to couple thermally excited surface waves into far-field thermal radiation. A selection of studies have also built on this concept to generate beamed thermal emission from 2D gratings \cite{Greffet2D,Metal2DHan:13,MiyazakiDualBand,Zhang2DThermophoto}. Another popular design for controlled thermal emission is the use of photonic crystals patterned in to metals or polar dielectrics \cite{SoljacicRinnerbauer:PhC}. 

The control of a material's epsilon-near-zero (ENZ) and optical topological transition characteristics has also proven to be a promising route for thermal emission engineering \cite{PendarkerThermGraph,MoleskyHighTENZ,MoleskyTopologicalTrans,GuoSuperPlanckian}. Plasma frequency tunable materials such as aluminum zinc oxide (AZO) or gallium zinc oxide (GZO) can provide a platform for implementation of this idea \cite{PendarkerThermGraph,MoleskyHighTENZ}. Intriguingly, a single thin layer of epsilon-near-zero material can function as a spectrally selective thermal emitter without the need for any nanostructuring. Multilayer thermal metamaterial designs have also experimentally demonstrated that thermal radiation can be controlled by engineering the optical topological transition \cite{MoleskyTopologicalTrans}. Such advances in thermal radiation control can impact a variety of commercial applications such as coherent infrared sources \cite{Greffet1D,Greffet2D,SchvetsIRemitterPolarMetasurf,SoljacicIncandescent}, thermophotovoltaics \cite{ZHANG1DGrating,MoleskyHighTENZ,WangTHermophoto,BermelTPV,Zhang2DThermophoto}, radiative cooling \cite{FanCooling,FanPhotonicStructures}, temperature sensors, gas sensors \cite{MiyazakiCO2}, nanoscale heat transfer \cite{EDALATPOURFrancoeurNearOverview,FrancoeurPhPThinFilm,ShenNearField,ShenNearMeta} and general thermal management.

Designs that exhibit dual band emission are also of interest and several studies towards this type of emission pattern have been published \cite{MiyazakiDualBand,PadillaBlackbody}. By designing emitters to have multi-band emission, we can increase functionality of the device for certain applications (e.g. sensing) by dedicating each emission band to perform a specific function. Despite the potential benefits of dual band emitters, all current designs have been limited to metallic structures and few have been experimentally demonstrated \citep{PadillaBlackbody,MiyazakiDualBand}. Furthermore, these demonstrations do not simultaneously control all emitter properties (spectrum, polarization, directionality, coherence) necessary for multiplexing of thermal signals. 

In this article, we design, fabricate and experimentally demonstrate a 2D bi-periodic SiC grating infrared source with orthogonally polarized, angularly coherent, dual band thermal emission in the mid-IR regime. We present a comprehensive experimental study of the thermal properties of this device by characterizing the spectral, polarization, angular, and temperature dependence of its thermal emission. The polarized emissivity spectrum is measured as both a function of temperature and emission angle within the Reststrahlen band of semi-insulating 6H-SiC (10.3-12.7~$\mu m$). Two emission bands, termed $\lambda_x$ and $\lambda_y$, are observed at normal incidence. This effect fundamentally arises due to thermally excited phonon polaritons in silicon carbide. The peak emissivity achieved for $\lambda_x$ and $\lambda_y$ are 0.85 and 0.81 respectively. The main emission peaks of each band shows an angular dependence, red shifting at higher emittance angles. A red shift of the emission peaks is also observed  with increasing temperatures. The peak of the $\lambda_x$ band shifts by 70 nm over a 255 $K$ temperature change, while the peak of the $\lambda_y$ band shifts by 60 nm over a temperature change of 259 $K$. Using a temperature dependent Drude model of SiC \cite{HERVETemp} in rigorous coupled wave analysis (RCWA) numerical simulations, strong agreement with our measurements is observed.


\section{Design, Fabrication and Characterization of Structure}

\subsection{Surface Phonon Polaritons in Polar Dielectrics}

Polar dielectric materials possess a spectral region known as the Reststrahlen band where a negative $Re(\varepsilon)$ (where $\varepsilon$ is the material permittivity) leads to metal-like behavior with high reflectivity. This region exists between the transverse $(TO)$ and longitudinal optical $(LO)$ phonons where polar lattice vibrations (phonon-polaritons) act to effectively screen electromagnetic waves, leading to high reflectivity analogous to a metal. Within the Reststrahlen band, coupled photonic-phononic surface waves, called surface phonon polaritons (SPhP), are supported, due to the interface between negative $Re(\varepsilon)$ and positive $Re(\varepsilon)$ materials \cite{CaldwellSPhP}. By attaching a piece of this material to a thermal bath (in our case a substrate heater), we can thermally populate these surface waves. Since the SPhP has a well defined dispersion relation, this provides a unique opportunity to engineer thermal emission in the Reststrahlen region in to well defined angular direction and spectral bands. SiC is one such polar dielectric that has been a popular choice for engineering thermal emission due to its high thermal stability (up to 3000~K) and its ability to support SPhP waves.

\subsection{Design of Bi-Periodic Grating}

Previous works have shown that the SPhP of SiC can be thermally excited and efficiently coupled to free space using 1D and 2D gratings \cite{Greffet2D,Greffet1D}. Here, we have designed and fabricated a bi-periodic grating to couple the thermally excited SPhP of SiC in to two bands of far-field propagating waves. The center wavelength of each emission peak can be independently tuned by modifying the x or y directions grating period as defined in figure 1 a). The structure is engineered such that each grating period ($\Lambda_x$ and $\Lambda_y$) will efficiently couple the SPhP in to a single band of emission ($\lambda_x=11.75~\mu m$ and $\lambda_y=12.25~\mu m$) that propagates with a $k$-vector normal to the surface ($\theta = 0$ as defined in figure 1 a)). Because surface waves are by nature $p$-polarized, the two engineered emission bands are orthogonally polarized in the far-field, providing an additional degree of discrimination.  

Figure 1 b) shows the real and imaginary permittivity of SiC, calculated using a Lorentz model \cite{Greffet2D} with the Reststrahlen band highlighted by the shaded region. On top of this we have plotted the measured emissivity of a bare SiC wafer at 1000 K (0.3 mm thick). Within the Reststrahlen band, low emissivity is indeed observed. This result supports Kirchoff's law as one would expect to see a high reflectivity here.

To calculate the emissivity spectrum in figure 1 b) (as well as emissivity spectra in later figures), the following approach was used from reference \cite{GreffetEmiss}:
\begin{equation}
E(\lambda)= \Big[ \frac{S_s(\lambda,T_s)-S_R(\lambda,T_R)}{S_B(\lambda,T_B)-S_R(\lambda,T_R)} \Big] \Big[ \frac{P(\lambda,T_B)-P(\lambda,T_R)}{P(\lambda,T_s)-P(\lambda,T_R)} \Big]
\end{equation}

where $S_s$ is the sample spectrum, $T_s$ is the temperature of the sample, $S_B$ is the spectrum of the blackbody calibration source, $T_B$ is the temperature of the blackbody calibration source, $S_R$ is the background spectrum at room temperature, $T_R$ is room temperature and $P(\lambda,T)$ is Planck's law at temperature $T$.

\begin{figure}[!htb]
\centering
\includegraphics[width=0.5\columnwidth]{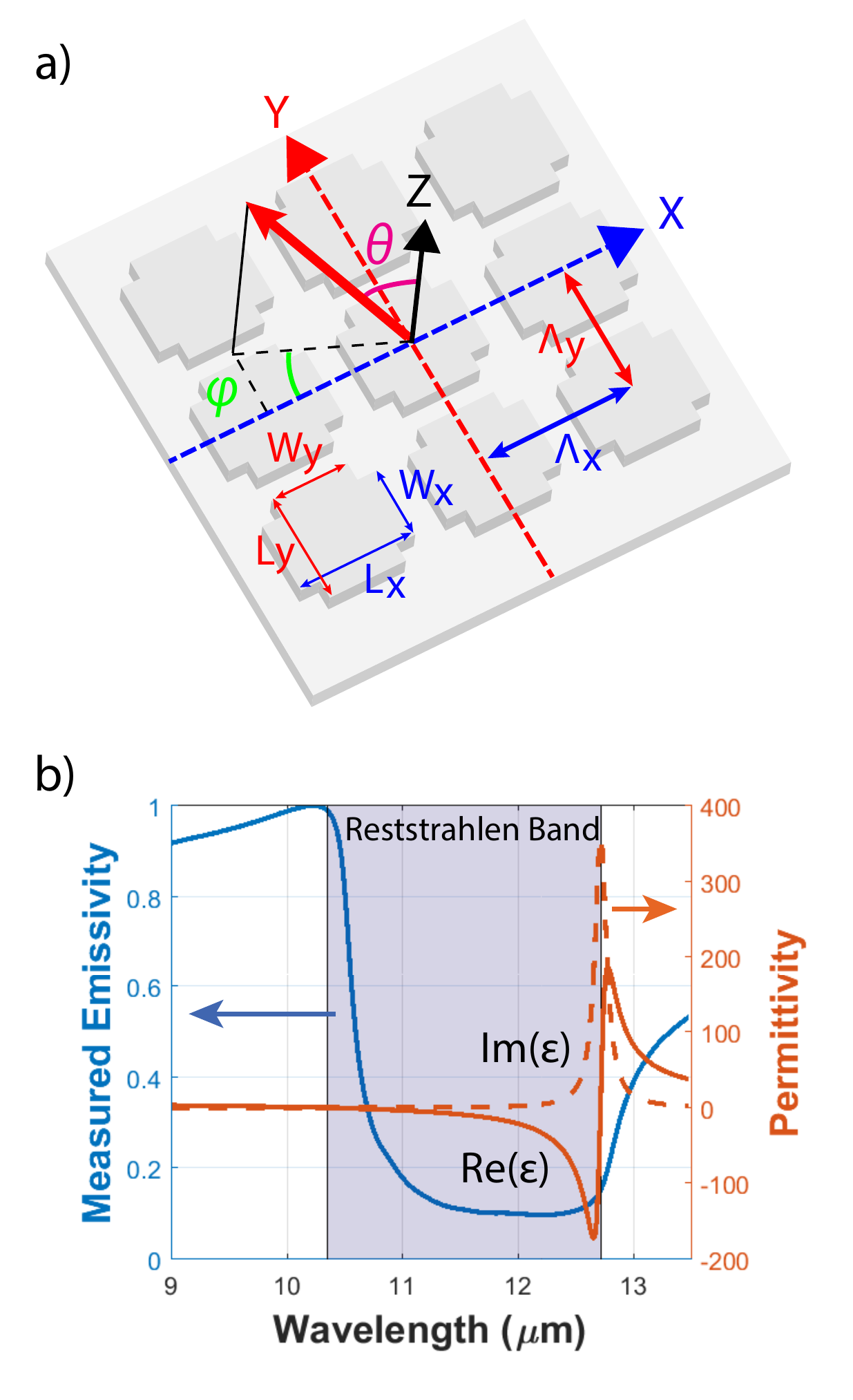}
\caption{a) Schematic of sample with axes, grating periods and emission angles defined. b) Real (solid red) and imaginary (dashed red) dielectric functions are plotted with the Reststrahlen band shaded. The blue curve shows measured emissivity at normal angle of emission of a bare SiC wafer. A low emissivity spectrum within the Reststrahlen band suggests high reflectivity, supporting the approximation $E(\lambda)$ = $A(\lambda)$ = $1-R(\lambda)$.}
\end{figure}

\subsection{Kirchoffs Law and Reflectivity/Emissivity Calculation}

According to Kirchoff's law of thermal radiation, a body that is in thermal equilibrium with its environment must have $E(\lambda,\theta) = A(\lambda,\theta)$ where $E(\lambda,\theta)$ is the emissivity spectrum and $A(\lambda,\theta)$ is the absorptivity spectrum of the material. It follows that if we are considering a material that is optically thick (transmission $= 0)$ we can calculate the emissivity of a material as \( E(\lambda,\theta) = 1-R(\lambda,\theta) \), where $R(\lambda,\theta)$ is the reflectivity spectrum. For typical polar materials, such as SiC, this condition is met for thicknesses greater than about 100 $\mu m$ \cite{Rousseau2005Christiansen}.

The permittivity function of a polar dielectric material is commonly modeled as a Lorentz oscillator \cite{Greffet1D}. However, it is known that at elevated temperatures the dielectric function of SiC changes. Here, we take in to account the changes in this material property at elevated temperatures by using a modified Drude model described by Herv\'{e} et al. \cite{HERVETemp}.

\begin{equation}
\varepsilon(\omega)=\varepsilon_\infty\Bigg[1-\frac{\Omega^2_p}{\omega(\omega+i\gamma_p)}\Bigg]+\frac{S\Omega^2}{\Omega^2-\omega^2-2\Omega \sum_{j=1}^{n} [\Delta_j(\omega)+i\Gamma_j(\omega)]}
\end{equation}

This model consists of two pieces: a high frequency contribution, $\varepsilon_\infty$ multiplied by a Drude term, with $\Omega_p$ acting as the plasma wavenumber, and $\gamma_p$ the Drude damping factor; and a phonon term \cite{SimonQuantumeps} with $S$ denoting the oscillator strength, $\Omega$ the wavenumber of the bare resonance frequency and $\sum[\Delta + i\Gamma]$ the self energy contribution. For simulations presented in this study, we have taken constants from ref. \cite{HERVETemp} at $996~K$ to compute the permittivity near our typical operating temperature.   

The two grating periods are designed such that the SPhP is coupled in to two directional emission bands almost normal to the surface. The two emission bands center wavelength of $\lambda_x=11.75~\mu m$ and $\lambda_y=12.25~\mu m$ are chosen so that they lie in the middle of the Reststrahlen band. We use RCWA simulations with the above dispersion relation to simulate the reflectivity of our design and determine the grating periods of $\Lambda_x=9.6~\mu m$ and $\Lambda_y=11.5~\mu m$ that are required to achieve the designed emission pattern. Furthermore, the dimensions $W_x$, $W_y$, $L_x$ and $L_y$ (as labeled in figure 1 a)) of the cross structure within the bi-periodic grating are optimized for efficient coupling of SPhP to achieve the maximum possible peak emissivities. 

\subsection{Fabrication and Characterization}

V-doped semi-insulating 6H-SiC wafers were purchased from Xiamen Powerway Advanced Material. Figure 2 a) shows a conventional lift off process flow used to fabricate the nickel mask for device fabrication. First, a 400 nm thick layer of PMMA A4 resist is spun on the SiC wafer and pre-baked at 180$^{\circ}$C for 90 s. Then, the resist layer is patterned by an electron-beam lithography (EBL) system and developed in a MIBK:IPA (1:3) solution for 1 min. Next, a 100 nm thick Ni metal film is deposited on the wafer using electron beam evaporation and a lift-off process is implemented using acetone as the solvent. The SiC wafer is then dry etched 700 nm deep using an SF6-based chemistry, creating a grating in the wafer. Finally, the Ni hard mask layer is removed by Piranha solution and the final structure is obtained. Figure 2 b) is an SEM of the final structure from a normal view.

\begin{figure}
\centering
\includegraphics[width=0.6\columnwidth]{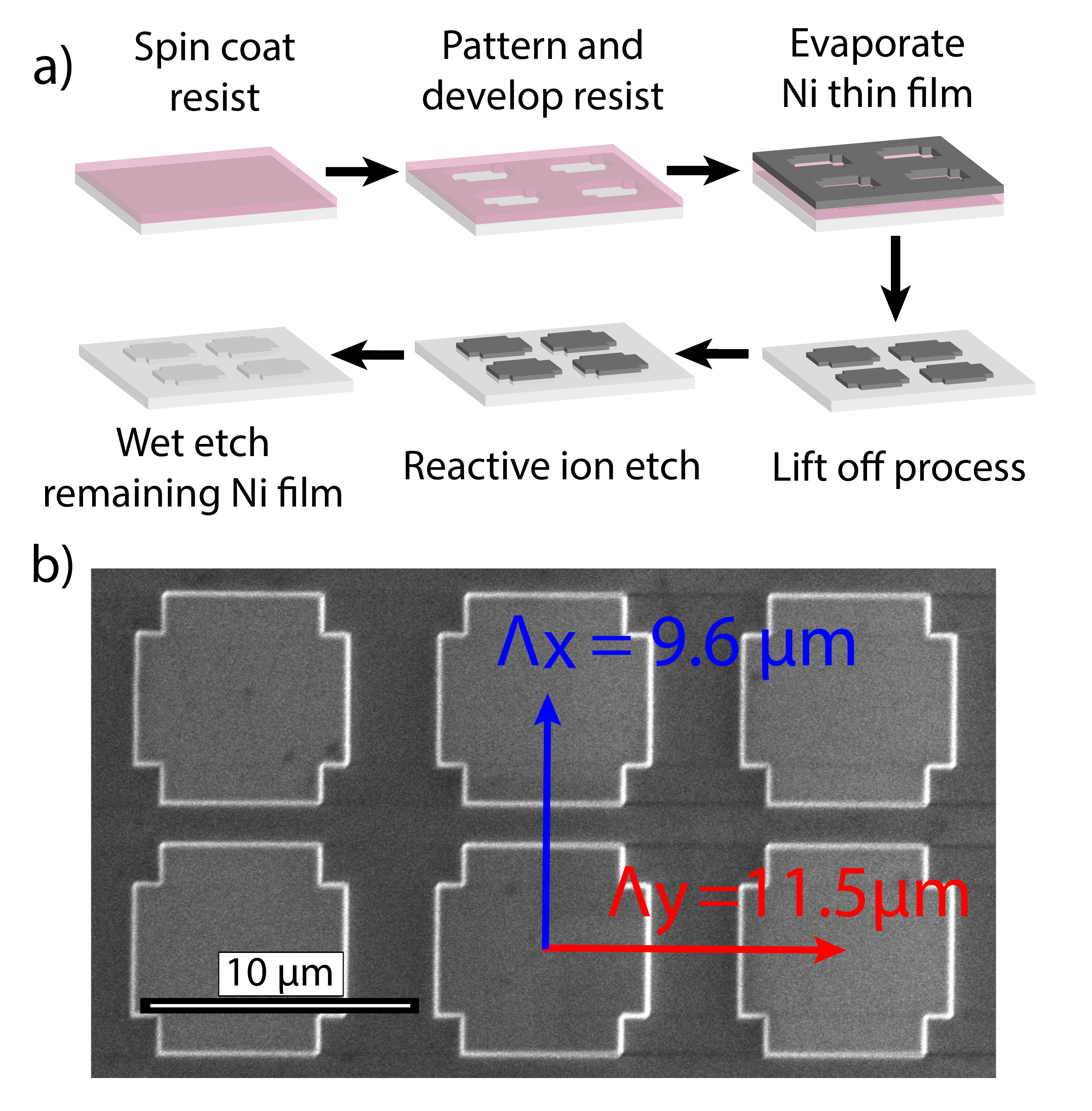}
\caption{a) Lift off process flow used to pattern bi-periodic grating in SiC. b) SEM image of final patterned structure with $\Lambda_x$ and $\Lambda_y$ periods labeled.}
\end{figure}


\section{Emissivity Measurements}

The emissivity spectrum of our structure is directly measured through thermal emission measurements. The signal is generated by controlled heating of the device on a substrate heater in vacuum. Detection is performed using a Thermo Scientific Nicolet is50 fourier transform infrared (FTIR) spectrometer. The measurement methods and results are described below.

\subsection{Experimental Thermal Emission Measurements and Results}

Thermal emission measurements are taken by collecting the temperature ($T$) and angular ($\theta$) dependent emission signal of the patterned SiC wafer. The emitter is attached to a high temperature substrate heater inside a custom built vacuum chamber and heated to temperatures from 689 to 963 K under vacuum (10$^{-7}$ mbarr). The emission signal is sent through a ZnSe window to allow for ex situ analysis and detection. Once the signal exits the chamber, its propagation path is confined to a nitrogen purged environment to reduce atmospheric absorption. The experimental set up is shown in figure 3. 

\begin{figure}[!htb]
\centering
\includegraphics[width=1\columnwidth]{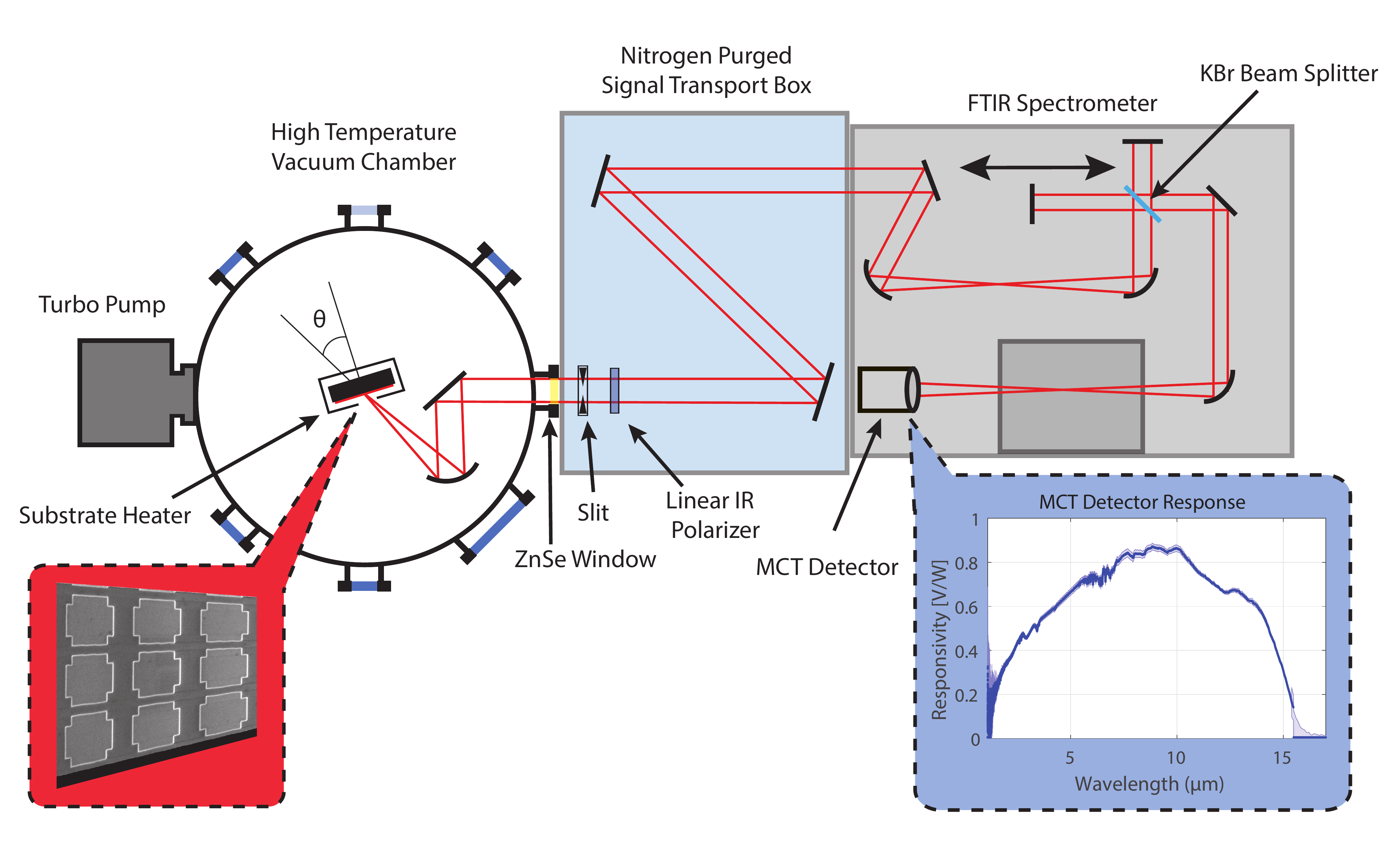}
\caption{Schematic of Experimental set up. Patterned sample is placed on the substrate heater inside the vacuum chamber. Emission signal is collected with a 4$"$ focal length parabolic mirror and is directed out of the chamber through a ZnSe window. A nitrogen purged signal transport box couples the signal in to the FTIR with proper alignment while housing optical analyzing elements (slits, polarizers etc.). For emissivity measurements from this structure, a MCT detector is typically used for high sensitivity in this spectral region.}
\end{figure}

After exiting the chamber, the signal passes through a slit to isolate the planar emission from one grating period ($\Lambda_x$ or $\Lambda_y$) at a time. The signal is then polarized using a KRS-5 wire grid linear polarizer. The combination of these two optical elements allows us to isolate $s$ or $p$ polarized light originating from either grating period of the structure. The slit is 22.86 mm long and 3 mm wide. With the collimating parabolic lens placed at its focal focal distance of 101.6~mm (4$"$), this corresponds to a collection angle of 12.8$^{\circ}$ and 1.7$^{\circ}$ respectively. For these measurements the FTIR spectrometer uses a KBr beam splitter and a mercury cadmium telluride (MCT) detector. The combined responsivity of the detection method spans the range of 1.4-16.7~$\mu m$ and is shown in figure 3. Measured spectra is averaged over 32 scans with a data spacing of 0.48 cm$^{-1}$.

To accurately determine the temperature of the emitter, we use the Christiansen wavelength method \cite{Rousseau2005Christiansen}. The Christiansen wavelength is a particular point in polar dielectric materials where there is a low extinction coefficient and the refractive index is 1. At this wavelength, there is a low impedance with free space so the reflectivity tends to 0. If the material is sufficiently thick such that the transmission is also 0 (as in this case), then it follows that absorptivity tends to 1. One can then justify, using Kirchoff's law of thermodynamics, that emissivity tends to 1 and at the Christiansen wavelength the material behaves like a black body. The emission of the device at this wavelength can be compared to Planck's law to accurately determine the temperature of operation. For SiC this occurs at 10.1 $\mu m$ \cite{Greffet2D}. 

Figure 4 shows the experimentally measured $p$-polarized emissivity of the SiC device at 963 K and $\theta = 0^{\circ}$ and its agreement with RCWA numerical simulations. Indeed we see two emission bands at $\lambda_x=11.75~\mu m$ and $\lambda_y=12.25~\mu m$. Of note is a second shoulder peak present in the emission spectrum of the $\lambda_y$ band. This shoulder peak results from the first order diffraction of the SPhP traveling in the opposite direction ($-\vec{k}_{SPhP}$). Figure 5 a) shows the temperature dependent $p$-polarized emission spectra resulting from the $\Lambda_x$ grating period ($\phi=0^\circ$). A shift of 70 nm in the peak wavelength is observed over a temperature change from 708 to 963 K. Figure 5 b) shows the temperature dependent $p$-polarized emission resulting from the $\Lambda_y$ grating period ($\phi=90^\circ$). A shift of 60 nm in the peak wavelength is observed for a temperature change from 689 to 948 K. This is consistent with previous studies and the temperature dependence of emission from SiC based grating structures is discussed in detail by Herv\'{e} et al. \cite{HERVETemp}. 

\begin{figure}[!htb]
\centering
\includegraphics[width=1\columnwidth]{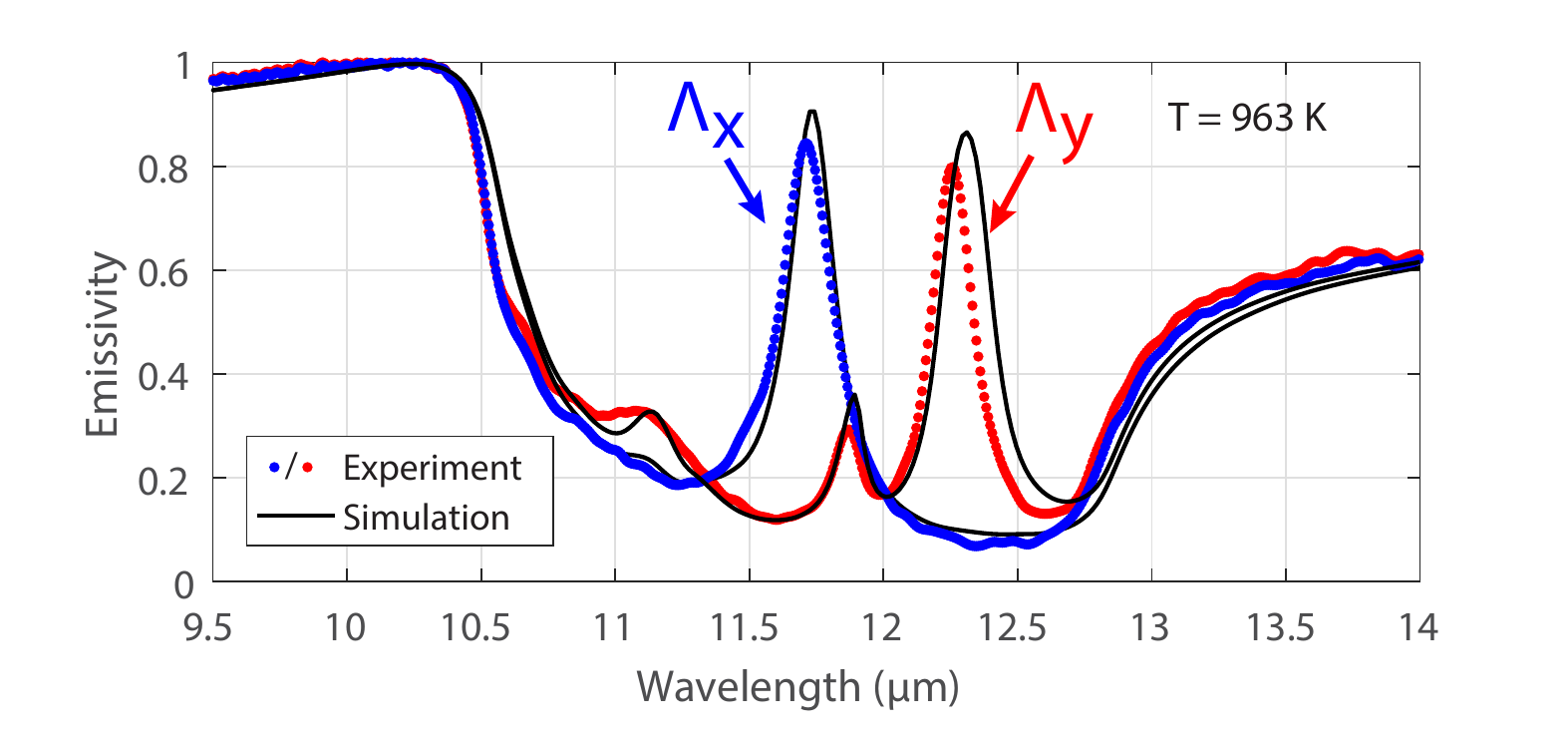}
\caption{Experimental (dot) vs simulated (line) $p$-polarized emissivity Spectrum of patterned SiC bi-periodic grating taken at normal angle of emission. The first peak located at 11.75 $\mu m$ is due to $\Lambda_x=9.6~\mu m$ while the second emission peak located at 12.25 $\mu m$ is due to $\Lambda_y = 11.5~\mu m$.}
\end{figure}

\begin{figure}[!htb]
\centering
\includegraphics[width=0.5\columnwidth]{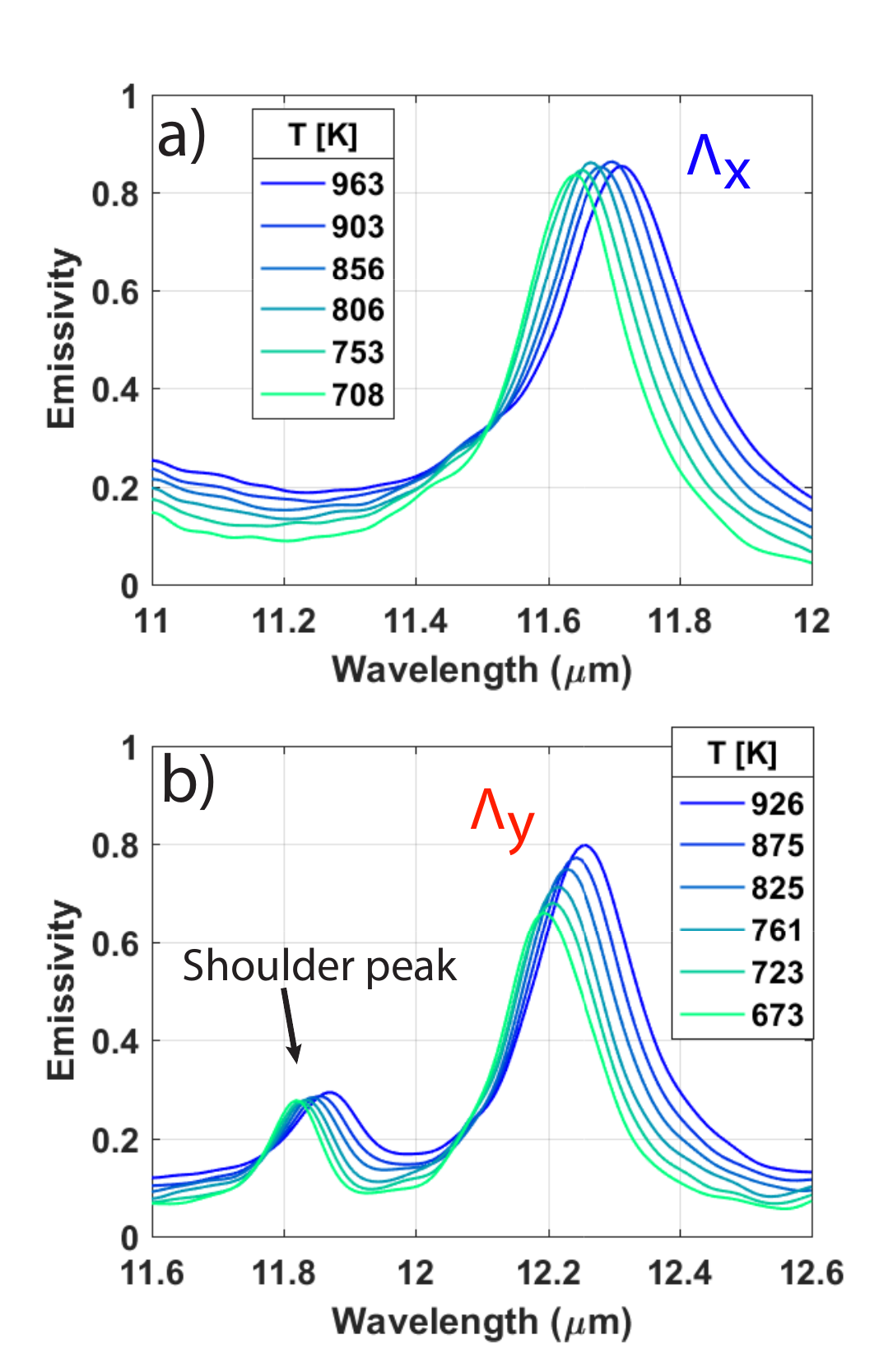}
\caption{a) Temperature dependent emission spectrum of the $p$-polarized $\lambda_x$ emission band. b) Temperature dependent emission spectrum of the $p$-polarized $\lambda_y$ emission band.}
\end{figure}

\subsection{Angularly Dependent Emission Spectra}

To further confirm the behavior of our bi-periodic grating design, angularly resolved numerical simulations are performed and compared to experiment. The simulated spectrum as a function of emission angle ($\theta$) is presented in figure 6 a)-d) where the 4 permutations of polarization ($s$ or $p$) and grating period ($\Lambda_x$ or $\Lambda_y$) are shown separately. We can see that for a normal angle of emission ($\theta = 0^{\circ}$) in figures 6 a) and b), there exists a high emissivity peak at 11.75 $\mu m$ and 12.25 $\mu m$ for $p$-polarized waves (the polarization for which this structure was designed to have high efficiency coupling between SPhP and free space propagating waves).

Figures 6 c) and d) shows that there exists a high emissivity peak for $s$-polarized waves. While our emitter is not designed to have such an emission pattern, it can be explained by out of plane scattering of SPhPs. Once a $p$-polarized surface wave is scattered to any wavevector that is not in the plane defined by the grating vector and the surface normal vector, a portion of it will then be $s$-polarized. This is common when coupling surface waves to free space and has been reported in previous studies \citep{Greffet2D,GreffetMarquier:OE}.

Figure 6 e)-h) shows the measured angularly resolved emissivity measurements. Measurements were taken at every 1$^{\circ}$ from 0$^{\circ}$-40$^{\circ}$ with an angular resolution of 1.7$^{\circ}$. For the two $p$-polarized cases figures 6 e) and f), we see an extremely good agreement between simulation and experiment. The main peak in the $p$-polarized emission is observed to shift to longer wavelengths at higher angles of emission ($\theta$). This can be attributed to the fact that the SPhP needs a larger angle ($\theta$) to be properly phase matched with free space propagating waves as described by the typical grating law:
\begin{equation}
\vec{k}_{SPhP}=\frac{\omega}{c}sin(\theta)-\frac{2 \pi m}{\Lambda}
\end{equation}

where $\vec{k}_{SPhP}$ is the SPhP wave vector, $m$ is the diffraction order integer ($m=\pm 1,\pm 2,...$), $\Lambda$ is the grating period, $\omega$ is the angular frequency and $\theta$ is the angle from the surface normal.

Additionally, there is good agreement between the experimental angular dependence of $s$-polarized emission and their theoretical simulations as shown in figures 6 g) and h). However, unlike the $p$-polarized case, the wavelength of this emission peak is not highly dependent on $\theta$. This is to be expected as the origin of this emission peak lies in the out of plane emission originating from the grating period orthogonal to the one being measured (grating period of $\phi=0$ when $\phi=90$ is being measured). For this reason, there is no $\theta$ dependence as described in equation (3).

Of note is some disagreement between simulation and experiment for the $s$-polarized case of $\Lambda_x$. This discrepancy can be attributed to the use of a slit for isolation of emission angles. In this particular case, the length of the slit is aligned along the Y period, allowing a large angular spread (12.8$^{\circ}$) of emission through the slit. This allows emission from the $\lambda_y$ shoulder peak to be detected as we can see it present in figure 6 g). This is not present in simulations as they are performed at precise angles and do not account for the finite window of angles being detected. Furthermore, because the main $p$-polarized $\lambda_y$ peak is highly sensitive to small changes at low $\theta$ angles, this leads to a reduced observed magnitude of the main peak (12.25 $\mu m$ at $\theta$=0$^{\circ}$) in figure 6 g). 

\begin{figure}[!htb]
\centering
\includegraphics[width=1\columnwidth]{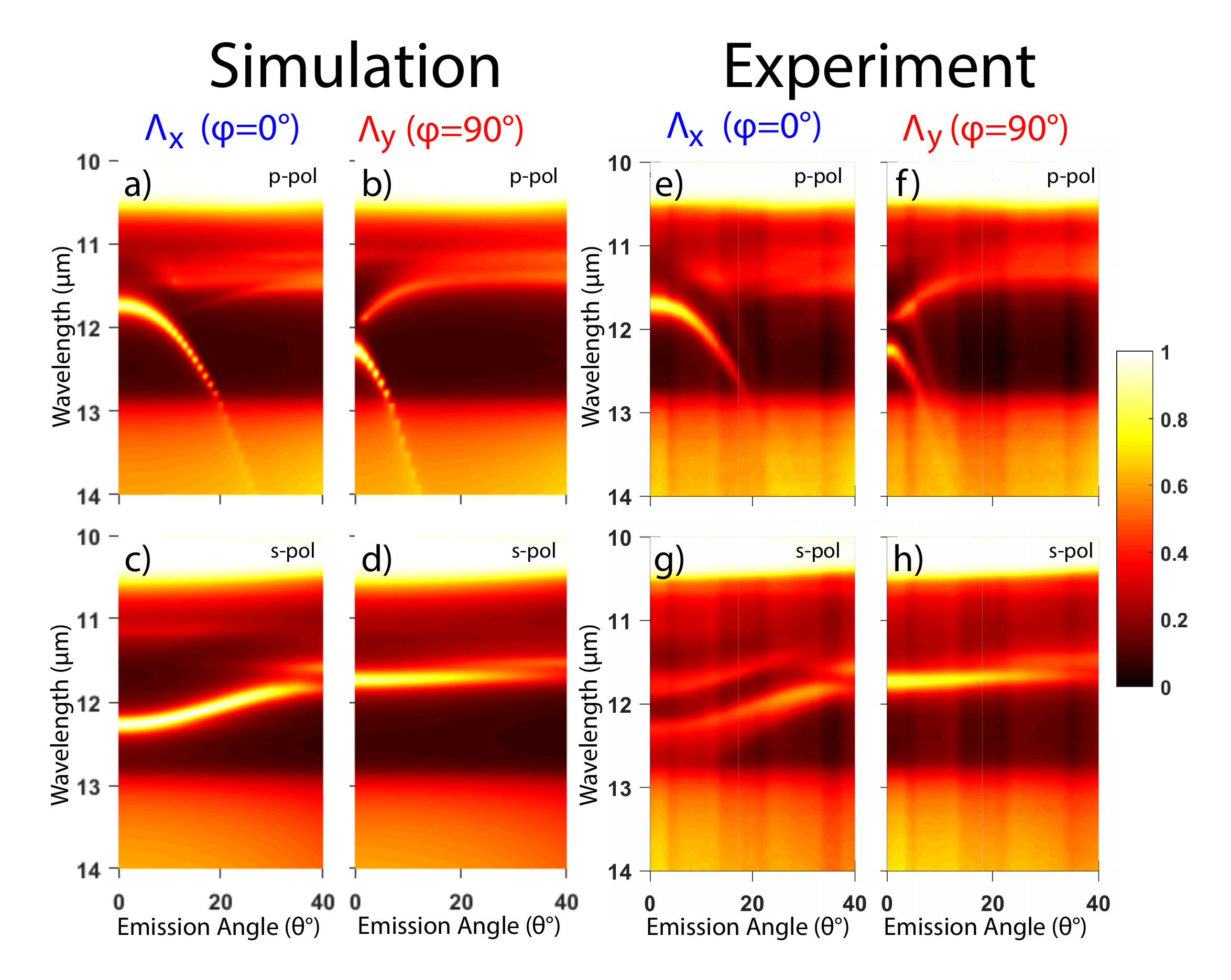}
\caption{a)-d) Simulated angular absorption/emission spectra of $p$-polarized (a) and b)) and $s$-polarized waves (c) and d)). Figures a) and c) are for emission that is propagating in the x-z plane ($\varphi = 0 ^{\circ}$) and figures b) and d) are for emission that is propagating in the y-z plane ($\varphi = 90 ^{\circ}$). RCWA numerical analysis is used to generate these simulations. e)-h) Experimental angular emissivity spectra of $p$-polarized (e) and f)) and $s$-polarized (g) and h)) waves. Figures e) and g) are for emission that is propagating in the x-z plane ($\varphi = 0 ^{\circ}$) and figures f) and h) are for emission that is propagating in the y-z plane ($\varphi = 90 ^{\circ}$).}
\end{figure}

\section{Conclusion}

We have successfully engineered a dual-band thermal emitter within the Reststrahlen region of semi-insulating SiC. The two spectral peaks are produced by coupling the SPhP in to free space propagating waves using a bi-periodic grating structure. We have fully characterized thermal emission from the device by analyzing its polarization, spectrum, angular emission pattern and temperature dependence. The center wavelength of the orthogonally polarized emission peaks can be independently controlled by adjusting the two grating periods. We have directly measured the the emissivity spectrum of the device through thermal emission measurements in vacuum and show that an emissivity of 0.85 is achieved for the $\lambda_x=11.75~\mu m$ band and 0.81 for the $\lambda_y=12.25~\mu m$ band. Our measurements are compared to RCWA numerical simulations and show they are in agreement when the temperature dependent material permittivity is accounted for. 

While this study reports on a SiC based device to exploit it's SPhP, the design principles can be easily transferred to metallic systems with surface plasmon-polaritons or other materials supporting surface waves.

\section*{Funding}

This material is based upon work supported by the National Science Foundation under Grant No.  EFMA-1641101


\bibliography{mybibfile}
\bibliographystyle{ieeetr}

\end{document}